\documentclass[floatfix,superscriptaddress,showpacs,amssymb,10pt,aps,prd,reprint,longbibliography]{revtex4-1}

\usepackage{graphicx,epsfig,amssymb}
\usepackage{amsmath,amsfonts, times}
\usepackage{bm}
\usepackage[linktocpage,colorlinks]{hyperref}
\usepackage[caption=false]{subfig}
\usepackage[usenames]{color}
\usepackage{natbib}
\usepackage[utf8x]{inputenc}
\usepackage{float}

\definecolor{darkgreen}{rgb}{0,0.5,0}
\definecolor{darkblue}{rgb}{0,0,0.5}
\hypersetup{colorlinks=true, citecolor=darkgreen, linkcolor=darkblue, urlcolor=blue}

\def\be{\begin{equation}}\def\ee{\end{equation}}

\newcommand{\bea}{\begin{eqnarray}}\newcommand{\eea}{\end{eqnarray}}

\begin{document}

\title{Thermodynamics and phase structure of rotating Hayward-AdS black holes}

\author{Qi-Quan Li}
\address{School of Physics Science and Technology, Xinjiang University, Urumqi 830046, China}

\author{Yu Zhang}
\email[Corresponding author:~]{zhangyu\_128@126.com}
\address{Faculty of Science, Kunming University of Science and Technology, Kunming 650500, China}

\author{Hoernisa Iminniyaz}
\email[Corresponding author:~]{wrns@xju.edu.cn}
\address{School of Physics Science and Technology, Xinjiang University, Urumqi 830046, China}

\begin{abstract}
We use the modified Newman--Janis algorithm to obtain the rotating ``mother'' black hole solution of the rotating Hayward-AdS black hole. We further derive the fundamental thermodynamic quantities of the ``mother'' black hole in the full unconstrained phase space $(S,J,q,\alpha,P)$, and impose the regularity condition $M_1=q^3/\alpha$ to obtain the thermodynamic quantities of the rotating Hayward-AdS regular black hole. This order of operations gives a Hawking temperature identical to the surface gravity temperature and produces the correct response functions of the regular black hole. For $P<P_c$, the Hawking temperature $T_2$ possesses a local maximum and a local minimum, which divides the black hole into small, intermediate, and large black hole branches, and the $P$--$V$ isotherms display oscillatory behavior characteristic of a liquid/gas phase transition. The small and large black hole branches are locally stable, whereas the intermediate black hole branch has negative heat capacity and is locally unstable. Increasing the rotation parameter $a$, magnetic charge $q$, or pressure $P$ reduces the region of the intermediate branch. The $G_2$--$T_2$ diagram shows that the phase structure, in contrast to the standard swallowtail of the Reissner--Nordstr\"{o}m-AdS and Kerr--Newman-AdS black holes, is 8-shaped or C-shaped. These two shapes are divided by the characteristic pressure $P_z$. For $P<P_z$, the curve is 8-shaped, corresponding to a first-order small/large black hole transition. For $P_z<P<P_c$, the curve is C-shaped, corresponding to a zeroth-order transition. These results show that the 8-shaped/C-shaped phase structure characteristic of regular AdS black holes persists under rotation, indicating that it stems from the restructuring of the thermodynamic phase space by the regularity condition rather than from spherical symmetry.
\end{abstract}

\maketitle

\section{Introduction}
\label{sec:1}

Since the 1970s, Hawking, Bekenstein, and Bardeen et al. established black hole thermodynamics~\cite{Bekenstein:1973ur,Bekenstein:1974jk,Hawking:1974rv,Hawking:1975vcx,Bardeen:1973gs}, the thermodynamic properties of black holes have been a popular research direction. However, the first law of black hole thermodynamics originally lacked a ``pressure--volume'' term. As a result, the heat capacity of a black hole is negative, and the black hole temperature rises during particle emission, so that it is an unstable thermodynamic system. It was pointed out that the negative cosmological constant in AdS spacetime and its conjugate quantity can be interpreted as the thermodynamic pressure and volume, respectively, thereby introducing a ``pressure--volume'' term into the first law~\cite{Creighton:1995au,Padmanabhan:2002sha,Teitelboim:1985dp,Brown:1987dd}.

In 2009, Kastor, Ray, and Traschen further systematized this idea by reinterpreting the black hole mass as an enthalpy~\cite{Kastor:2009}. The $P$--$V$ criticality in the extended phase space was discovered, and AdS black holes possessed stable phases. Later, the $P$--$V$ criticality and the first-order small/large black hole phase transition of Reissner--Nordstr\"{o}m-AdS (RN-AdS) black holes were revealed~\cite{Kubiznak:2012,Gunasekaran:2012}, analogous to the liquid/gas transition of a van der Waals fluid, whose signature is the swallowtail structure of the Gibbs free energy~\cite{Chamblin:1999tk,Chamblin:1999hg}. For the rotating Kerr--Newman-AdS (KN-AdS) black holes, a complete thermodynamic description was provided via a generalized Smarr formula~\cite{Caldarelli:2000nba}, expressing the mass as a fundamental relation $M(S,J,Q)$ whose partial derivatives reproduce the temperature, angular velocity, and electric potential. These advances have provided a rich and self-consistent framework for the singular AdS black holes, and the thermodynamics of rotating AdS black holes has by now become a relatively mature research direction.

However, this self-consistency does not automatically extend to regular black holes, which are constructed without the central curvature singularity. Regular black holes can be constructed in various ways, for example by coupling Einstein gravity to nonlinear electrodynamics, which produces exact solutions such as the Bardeen~\cite{Bardeen:1968,Ayon-Beato:2000mjt}, Ay\'{o}n-Beato--Garc\'{i}a~\cite{Ayon-Beato:1998hmi}, Hayward~\cite{Hayward:2005gi}, and Fan--Wang~\cite{Fan:2016hvf,Toshmatov:2018cks} black holes, in which the mass function is designed to be zero at the origin, so that no curvature singularity exists. However, for such regular black holes, the Hawking temperature computed from surface gravity differed from the temperature derived from the first law of thermodynamics~\cite{Ma:2014qma} for a long time. It was not until 2025 that Ma et al.~\cite{Ma:2025reduction} traced this inconsistency to the regularity condition $\lim\limits_{r\to0}m(r)=0$, which imposes a constraint among the black hole parameters, reduces the dimension of the thermodynamic phase space, and distorts the first law of black hole thermodynamics when it is imposed prematurely. The systematic resolution is to first derive the fundamental thermodynamic quantities in the full unconstrained phase space of a singular ``mother'' black hole, where the first law takes its standard form and the two temperatures coincide, and then to impose the constraint on the thermodynamic quantities of the ``mother'' black hole, thereby obtaining the fundamental thermodynamic quantities of the regular black hole.

In this paper, we extend the above approach to the rotating Hayward-AdS black hole. A rotating black hole in AdS space differs from its static counterpart in that the AdS boundary itself rotates. The physical mass is related to the metric parameter by $M_{\rm phys} = M_{\rm met}/\Xi^2$ with $\Xi = 1 - 8\pi P a^2/3$, and the angular velocity is measured relative to the rotating boundary, $\Omega = \Omega_H - \Omega_\infty$~\cite{Caldarelli:2000nba,Gibbons:2004ai}. We obtain the entropy and the angular velocity of the ``mother'' black hole from its geometry, and the remaining fundamental thermodynamic quantities from the standard first law in the unconstrained phase space. Then the regularity condition is imposed on these fundamental thermodynamic quantities to obtain the thermodynamic quantities of the rotating Hayward-AdS black hole. On this basis, we further present a systematic analysis of the phase structure of the rotating Hayward-AdS black hole.

The analysis shows that the phase structure is governed by two characteristic pressures, the critical pressure $P_c$ and a characteristic pressure $P_z$. For $P<P_c$, the Hawking temperature $T_2$ possesses a local maximum and a local minimum, which divides the black hole into the small black hole (SBH), intermediate black hole (IBH), and large black hole (LBH) branches. The intermediate branch is locally unstable, whereas the small and large branches are locally stable. The Gibbs free energy does not display the standard swallowtail of the RN-AdS and KN-AdS black holes. For $P<P_z$, the $G_2$--$T_2$ curve is 8-shaped, corresponding to a first-order small/large black hole transition, whereas for $P_z<P<P_c$ it is C-shaped, corresponding to a zeroth-order transition. At $P=P_c$, a second-order phase transition occurs, and for $P>P_c$ no phase transition exists.

This paper is organized as follows. In Sec.~\ref{sec:2}, we apply the modified Newman--Janis algorithm to the spherically symmetric ``mother'' black hole of the Hayward-AdS black hole to construct its rotating counterpart. In Sec.~\ref{sec:3}, we derive the thermodynamic quantities of the Hayward-AdS black hole. In Sec.~\ref{sec:4}, we study the phase structure, local stability, and $P$--$V$ criticality of the rotating Hayward-AdS black hole. In Sec.~\ref{sec:5}, we summarize and discuss the results of this paper.
\section{The rotating ``mother'' black hole of the Hayward-AdS black hole}
\label{sec:2}
\begin{figure*}
\centering
\subfloat[][]{\includegraphics[scale=0.47]{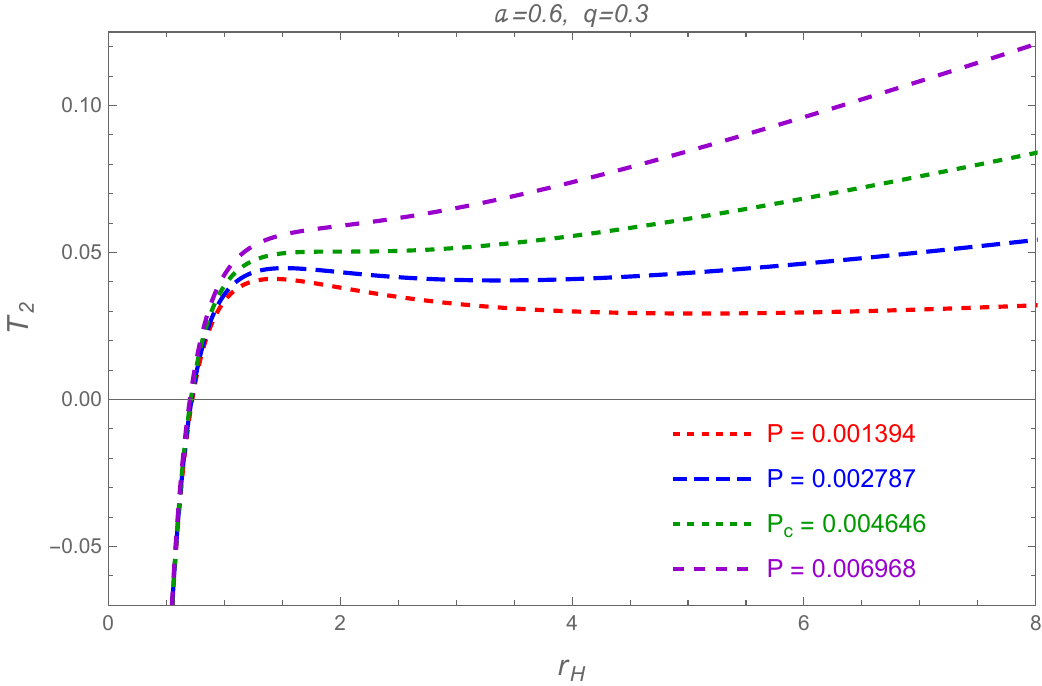}}\hspace{0cm}
\subfloat[][]{\includegraphics[scale=0.47]{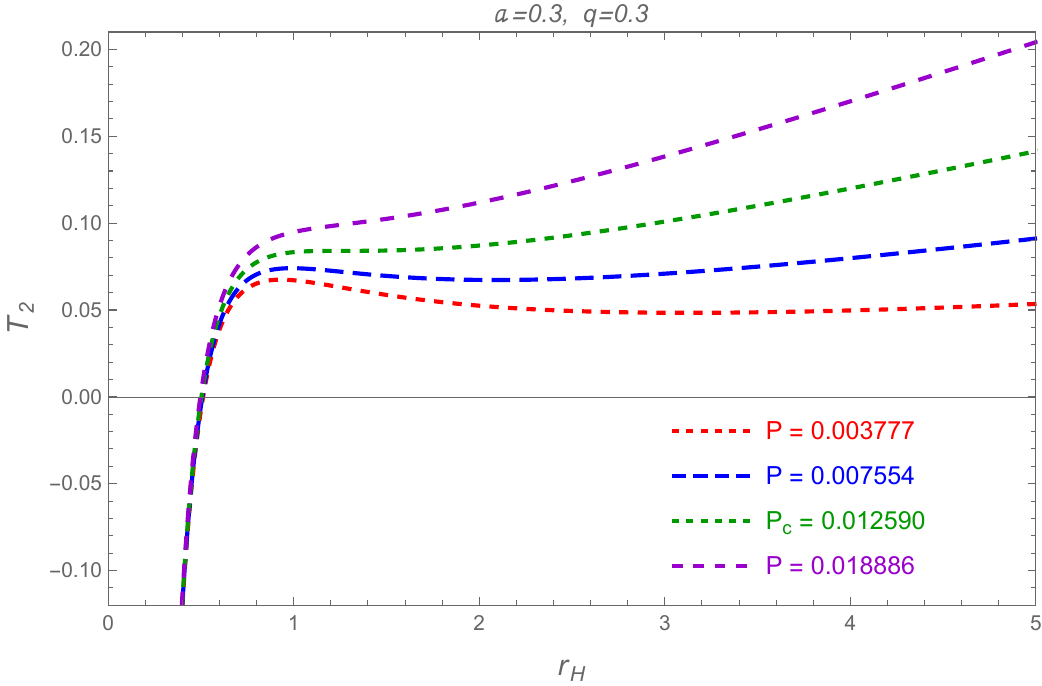}}\\[2mm]
\caption{The behavior of $T_2$ vs $r_H$ for different parameters $a$, $q$ and $P$.}
\label{fig:Tr}
\end{figure*}

In this section, we obtain the ``mother'' black hole of the rotating Hayward-AdS black hole from its spherically symmetric counterpart. In Ref.~\cite{Li:2026ewf}, we studied the thermodynamics and phase transitions of a class of spherically symmetric AdS regular black holes arising from Einstein gravity coupled to nonlinear electrodynamics, and obtained the mass function of the ``mother'' black hole for this class of regular black holes. The construction there begins with the nonlinear electrodynamics Lagrangian, and the mass function takes the form~\cite{Fan:2016hvf,Toshmatov:2018cks}
\begin{equation}\label{eq:22}
m(r) = M - \frac{q^{3}}{\alpha}\Bigl(1 - \frac{r^{\mu}}{(r^{\nu} + q^{\nu})^{\mu/\nu}}\Bigr) + \frac{\Lambda}{6}r^{3},
\end{equation}
where the parameters $M$, $q$, and $\alpha$ are independent. Here $M$ is the ADM mass parameter. The regularization parameter $q$ is related to the magnetic charge $Q_m$ of the nonlinear electromagnetic source through $Q_m = q^{2}/\sqrt{2\alpha}$, and $\alpha>0$ is the coupling constant of the nonlinear electrodynamics, with dimensions of length squared. For $\mu=\nu=3$, the mass function reduces to that of the ``mother'' black hole of the Hayward-AdS black hole,
\begin{equation}\label{eq:22b}
m(r) = M - \frac{q^{6}}{\alpha(q^{3}+r^{3})} + \frac{\Lambda}{6}r^{3}.
\end{equation}
From this mass function, we obtain the static spherically symmetric line element of the ``mother'' black hole of the Hayward-AdS black hole as
\begin{equation}\label{eq:22c}
ds^{2} = -f(r)dt^{2} + f(r)^{-1}dr^{2} + r^{2}(d\theta^{2} + \sin^{2}\theta\,d\phi^{2})
\end{equation}
where
\begin{equation}\label{eq:22d}f(r) = 1 - \frac{2M}{r} + \frac{2q^{6}}{\alpha r(q^{3}+r^{3})} - \frac{\Lambda}{3}r^{2}.
\end{equation}
When the regularity condition $M = q^{3}/\alpha$~\cite{Fan:2016hvf,Li:2025spherical} is imposed, this solution reduces to the spherically symmetric Hayward-AdS black hole.

Since the Newman--Janis algorithm was proposed in 1965~\cite{Newman:1965tw}, it has been widely applied to construct rotating solutions from static spherically symmetric black holes~\cite{Benavides-Gallego:2018odl,Liu:2020ola,Xu:2020jpv,Kumar:2017qws,Xu:2016jod,Mazza:2021rgq,Bambi:2013ufa}. Beginning with the spherically symmetric ``mother'' black hole of the Hayward-AdS black hole and applying the Newman--Janis algorithm in the modified form of Ref.~\cite{Azreg-Ainou:2014pra}, we obtain the expression for its rotating ``mother'' black hole, which in Boyer--Lindquist coordinates reads
\begin{equation}\label{eq:24}
\begin{aligned}
ds^{2} =& -\frac{\Delta_r}{\Sigma}\Bigl(dt - \frac{a\sin^{2}\theta}{\Xi}d\phi\Bigr)^{2}
        + \frac{\Sigma}{\Delta_r}dr^{2} + \frac{\Sigma}{\Delta_\theta}d\theta^{2}\\
       &+ \frac{\Delta_\theta\sin^{2}\theta}{\Sigma}\Bigl(a\,dt - \frac{r^{2}+a^{2}}{\Xi}d\phi\Bigr)^{2},
\end{aligned}
\end{equation}
where
\begin{equation}\label{eq:25}
\Sigma = r^{2} + a^{2}\cos^{2}\theta,\qquad
\Xi = 1 + \frac{\Lambda a^{2}}{3},
\end{equation}
\begin{gather}
\Delta_\theta = 1 + \frac{\Lambda a^{2}}{3}\cos^{2}\theta,\\[2mm]
\Delta_r = (r^{2}+a^{2})(1 - \frac{\Lambda r^{2}}{3}) - 2r(M - \frac{q^{6}}{\alpha(q^{3}+r^{3})}).\label{eq:25b}
\end{gather}
Substituting the regularity condition $M = q^{3}/\alpha$ into the above line element, it recovers the rotating Hayward-AdS black hole solution.

\section{Thermodynamics of the rotating AdS regular  black hole}
\label{sec:3}

In this section, we first present the fundamental thermodynamic quantities of the singular ``mother'' black hole in the thermodynamic phase space $(S, J, q, \alpha, P)$, and then impose the regularity condition $M = q^{3}/\alpha$ on them, thereby obtaining the fundamental thermodynamic quantities of the rotating regular black hole.

\subsection{Thermodynamics of the rotating ``mother'' black hole}

In this subsection, we derive the fundamental thermodynamic quantities of the singular ``mother'' black hole in the full unconstrained phase space $(S, J, q, \alpha, P)$. We consider the extended phase space, in which the cosmological constant is traded for the thermodynamic pressure $P = -\Lambda/(8\pi)$. Solving the condition $\Delta_r(r_H)=0$, the black hole mass is
\begin{equation}\label{eq:mother}
M_1 = \frac{(r_H^{2}+a^{2})\bigl(1 + \frac{8\pi P}{3}r_H^{2}\bigr)}{2r_H}
      + \frac{q^{6}}{\alpha(q^{3}+r_H^{3})},
\end{equation}
where $r_H$ is the event horizon radius. The above equation $M_1$ is only the mass parameter appearing in the metric, and it is not the conserved mass of the spacetime. For a rotating AdS black hole, the conserved mass defined by the boundary charges at infinity differs from the metric parameter and is given by $M_{\rm phys} = M_1/\Xi^{2}$~\cite{Gibbons:2004ai}, with $\Xi = 1 - \frac{8\pi P a^{2}}{3}$. In the extended thermodynamic phase space, this physical mass is not the internal energy but the enthalpy.

According to the Bekenstein--Hawking area law, the entropy of the rotating ``mother'' black hole is
\begin{equation}\label{eq:31}
S = \frac{A}{4}
= \frac{1}{4}\int_0^{2\pi}\int_0^{\pi}\sqrt{g_{\theta\theta}\,g_{\varphi\varphi}}\,d\theta\,d\varphi
= \frac{\pi(r_H^{2}+a^{2})}{\Xi}.
\end{equation}
The angular momentum is $J = aM_{\rm phys}$, and the angular velocity at the event horizon relative to the rotating AdS boundary is obtained as
\begin{equation}\label{eq:32}
\Omega = \frac{a\bigl(1 + \frac{8\pi P}{3}r_H^{2}\bigr)}{r_H^{2}+a^{2}}.
\end{equation}

For the ``mother'' black hole of the rotating Hayward-AdS black hole, the first law of thermodynamics in the unconstrained phase space $(S, J, q, \alpha, P)$ reads
\begin{equation}\label{eq:35}
dM_{\rm phys} = T_1 dS + \Omega dJ + \Phi_1 dq + \mathcal{A}_1 d\alpha + V_1 dP,
\end{equation}
where the subscript ``1'' indicates that these quantities refer to the singular ``mother'' black hole. Here $T_1$, $\Omega$, $\Phi_1$, $\mathcal{A}_1$, and $V_1$ are the conjugate variables of $S$, $J$, $q$, $\alpha$, and $P$, respectively. In Eq.~(\ref{eq:35}), the angular velocity $\Omega$ does not involve the parameter $\alpha$ and is therefore unaffected by the regularity condition. It is identical for the ``mother'' and the regular black hole and thus carries no subscript.
According to the first law of black hole thermodynamics Eq.~(\ref{eq:35}) and the definitions of the thermodynamic quantities, the conjugate thermodynamic quantities of the ``mother'' black hole of the rotating Hayward-AdS black hole are obtained as
\begin{equation}\label{eq:36}
\begin{aligned}
T_1 &\equiv \left(\frac{\partial M_{\rm phys}}{\partial S}\right)_{J,q,\alpha,P}\\
&= \frac{\alpha(q^{3}+r_H^{3})^{2}\bigl(r_H^{2} - a^{2} + \frac{8\pi P}{3}r_H^{2}(a^{2}+3r_H^{2})\bigr) - 6q^{6} r_H^{4}}
        {4\pi r_H\,\alpha\,(q^{3}+r_H^{3})^{2}(r_H^{2}+a^{2})}.
\end{aligned}
\end{equation}
\begin{equation}\label{eq:37}
\Phi_1 \equiv \left(\frac{\partial M_{\rm phys}}{\partial q}\right)_{S,J,\alpha,P}
       = \frac{3q^{5}(q^{3}+2r_H^{3})}{\alpha\,\Xi^{2}\,(q^{3}+r_H^{3})^{2}}.
\end{equation}
\begin{equation}\label{eq:37a}
\mathcal{A}_1 \equiv \left(\frac{\partial M_{\rm phys}}{\partial \alpha}\right)_{S,J,q,P}
              = -\frac{q^{6}}{\alpha^{2}\,\Xi^{2}\,(q^{3}+r_H^{3})}.
\end{equation}
\begin{equation}\label{eq:37b}
\begin{aligned}
V_1 &\equiv \left(\frac{\partial M_{\rm phys}}{\partial P}\right)_{S,J,q,\alpha}\\
&= \frac{4\pi a^{2} q^{6} r_H^{3}}{\alpha\,\Xi^{2}(q^{3}+r_H^{3})^{2}} + \frac{16\pi a^{2} M_1 r_H^{2}}{3\,\Xi^{2}(r_H^{2}+a^{2})}\\
&\quad - \frac{2\pi a^{2}\bigl(r_H^{2}-a^{2}+\frac{8\pi P}{3}r_H^{2}(a^{2}+3r_H^{2})\bigr)}{3\,\Xi^{2} r_H} + \frac{4\pi r_H^{3}}{3\,\Xi}.
\end{aligned}
\end{equation}
One can verify that these thermodynamic quantities satisfy the Smarr relation
\begin{equation}\label{eq:37c}
\begin{aligned}
M_{\rm phys} = 2T_1 S + 2\Omega J + \Phi_1 q + 2\mathcal{A}_1 \alpha - 2P V_1.
\end{aligned}
\end{equation}

\subsection{Thermodynamics of the rotating Hayward-AdS black hole}

\begin{figure*}
\centering
\subfloat[][]{\includegraphics[scale=0.47]{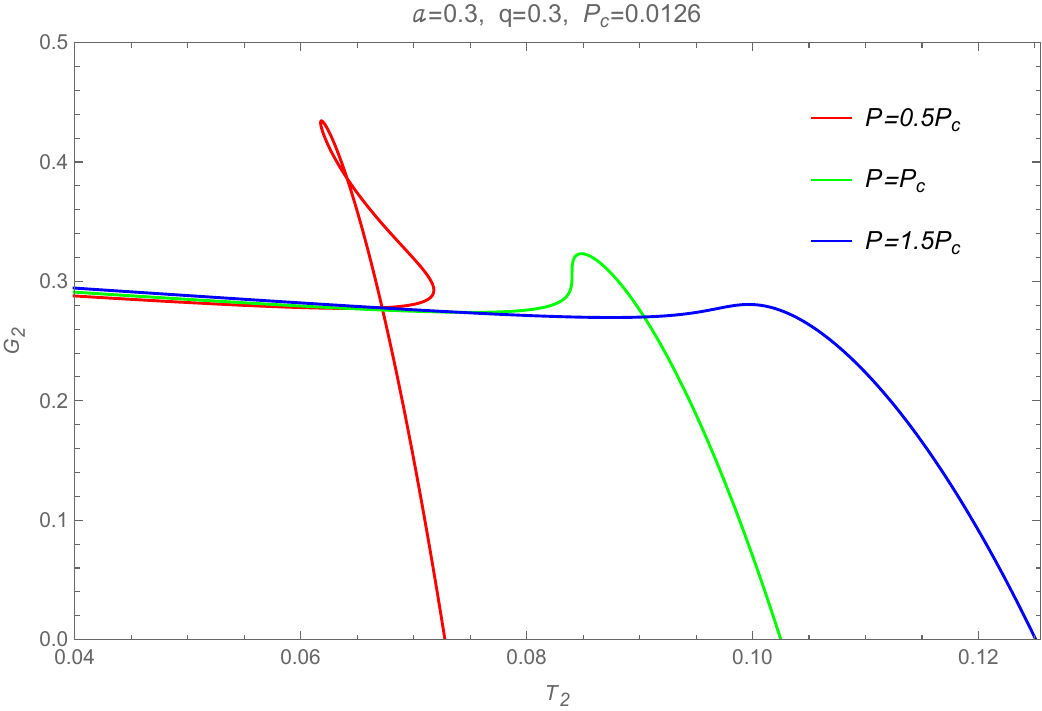}}\hspace{0cm}
\subfloat[][]{\includegraphics[scale=0.47]{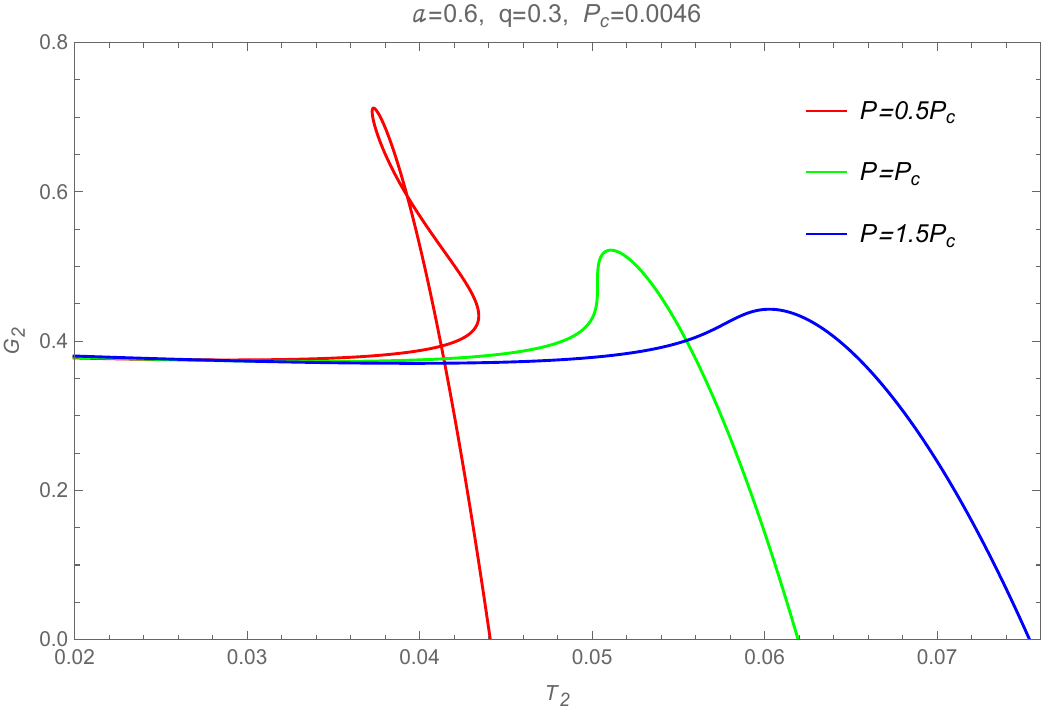}}\\[2mm]
\caption{The behavior of $G_2$ vs $T_2$ for different parameters $a$, $q$ and $P$.}
\label{fig:GT_3curves}
\end{figure*}
We now impose the regularity condition $M = q^{3}/\alpha$ on the thermodynamic quantities of the ``mother'' black hole obtained in the previous subsection, thereby deriving the thermodynamic quantities of the rotating Hayward-AdS black hole. Substituting the regularity condition into the mass $M_1$, we obtain
\begin{equation}\label{eq:38}
M_2 = \frac{(r_H^{2}+a^{2})(3 + 8\pi P r_H^{2})(q^{3}+r_H^{3})}{6r_H^{4}}.
\end{equation}
We substitute the above equation and the regularity condition into the thermodynamic quantities of the ``mother'' black hole, obtaining the thermodynamic quantities of the rotating Hayward-AdS black hole as
\begin{equation}\label{eq:39}
\begin{aligned}
T_2 &= \frac{1}{12\pi r_H(r_H^{2}+a^{2})(q^{3}+r_H^{3})}
      \Bigl[3\bigl(-2q^{3}r_H^{2}+r_H^{5}+8\pi P r_H^{7}\bigr)\\
&\qquad - a^{2}\bigl(3r_H^{3}-8\pi P r_H^{5}+4q^{3}(3+4\pi P r_H^{2})\bigr)\Bigr],
\end{aligned}
\end{equation}
\begin{equation}\label{eq:310}
\Phi_2 = \frac{3q^{2}(r_H^{2}+a^{2})(1+\frac{8\pi P}{3}r_H^{2})}{\Xi^{2}\,r_H^{4}}
         \Bigl[1 - \frac{q^{3}}{2(q^{3}+r_H^{3})}\Bigr],
\end{equation}
\begin{equation}\label{eq:311}
\mathcal{A}_2 = -\frac{(r_H^{2}+a^{2})^{2}(1+\frac{8\pi P}{3}r_H^{2})^{2}(q^{3}+r_H^{3})}
                       {4\,\Xi^{2}\,r_H^{8}},
\end{equation}
\begin{equation}\label{eq:312}
\begin{aligned}
V_2 = &-\frac{2\pi a^{2}}{9\,\Xi^{2}\, r_H\,(q^{3}+r_H^{3})}
      \Bigl[3\bigl(-2q^{3}r_H^{2}+r_H^{5}+8\pi P r_H^{7}\bigr)\\
&\quad - a^{2}\bigl(3r_H^{3}-8\pi P r_H^{5}+4q^{3}(3+4\pi P r_H^{2})\bigr)\Bigr]\\
&+ \frac{4\pi r_H^{3}}{3\,\Xi} + \frac{16\pi a^{2} M_2 r_H^{2}}{3\,\Xi^{2}(r_H^{2}+a^{2})},
\end{aligned}
\end{equation}
where the subscript ``2'' indicates that these quantities refer to the rotating Hayward-AdS black hole. It is straightforward to verify that $T_2$ coincides with the surface gravity temperature of the constrained metric.

Now we specifically investigate the properties of the Hawking temperature $T_2$. The behavior of $T_2$ with respect to the event horizon radius $r_H$ for various values of the pressure $P$ is shown in Fig.~\ref{fig:Tr}. When the local maximum and minimum of $T_2$ coincide, the temperature satisfies
\begin{equation}\label{eq:Tcrit}
\frac{\partial T_2}{\partial r_H}=0 \quad \text{and} \quad \frac{\partial^2 T_2}{\partial r_H^2}=0.
\end{equation}
From Eq.~(\ref{eq:Tcrit}), for given values of $a$ and $q$, the critical pressure $P_c$ can be obtained numerically. By analyzing Eq.~(\ref{eq:39}) and Fig.~\ref{fig:Tr}, we get the following conclusions:
\begin{itemize}
\item When $P<P_c$, $T_2$ first increases with $r_H$ to a local maximum at $r_H=r_{H1}$, then decreases to a local minimum at $r_H=r_{H2}$, and finally increases again. These two turning points partition the black hole into three branches, namely the SBH branch for $0<r_H<r_{H1}$, the IBH branch for $r_{H1}<r_H<r_{H2}$, and the LBH branch for $r_H>r_{H2}$.
\item When $P=P_c$, the two stationary points coalesce into a single inflection point, and the IBH branch disappears. For $P>P_c$, $T_2$ increases monotonically with $r_H$, so that no phase transition can occur.
\end{itemize}
As discussed above, the two stationary points of the Hawking temperature and the non-monotonic behavior of $T_2$ are related to, respectively, the local thermodynamic stability and the phase transition features of the $G_2$--$T_2$ diagrams to be analyzed in the next section.

\section{Phase transitions and thermodynamic stability}
\label{sec:4}
\begin{figure*}
\centering
\subfloat[][]{\includegraphics[scale=0.47]{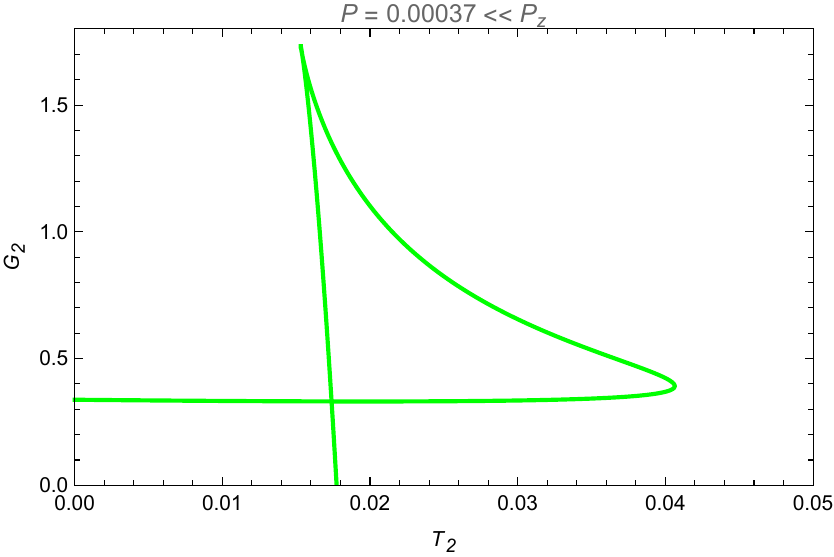}}\hspace{0cm}
\subfloat[][]{\includegraphics[scale=0.47]{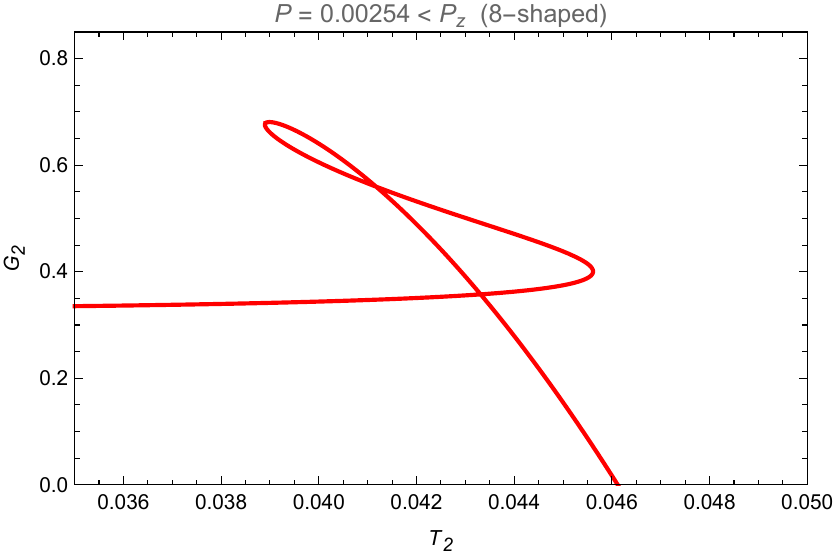}}\\[2mm]
\subfloat[][]{\includegraphics[scale=0.47]{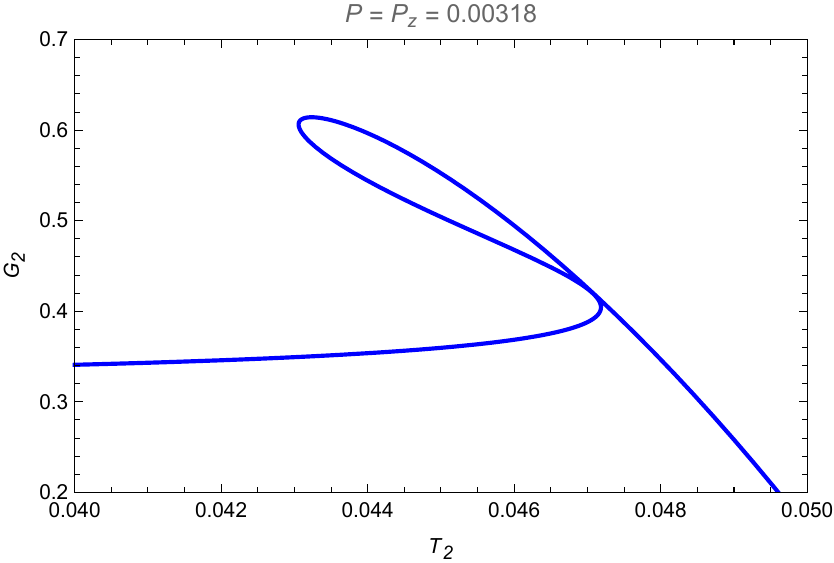}}\hspace{0cm}
\subfloat[][]{\includegraphics[scale=0.47]{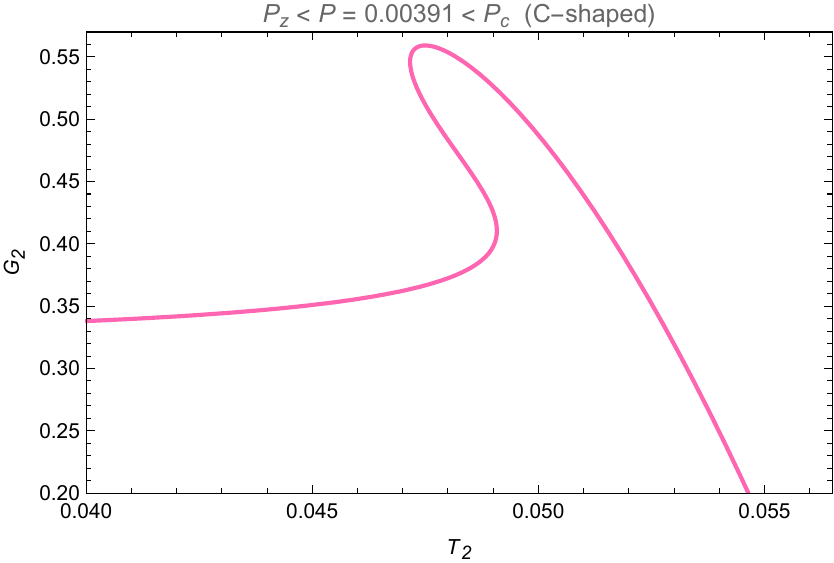}}
\caption{The behavior of $G_2$ vs $T_2$ for the rotating Hayward-AdS black hole with $a=0.6$, $q=0.3$ and different values of $P$.}
\label{fig:GT_4panels}
\end{figure*}
\begin{figure*}
\centering
\subfloat[][]{\includegraphics[scale=0.47]{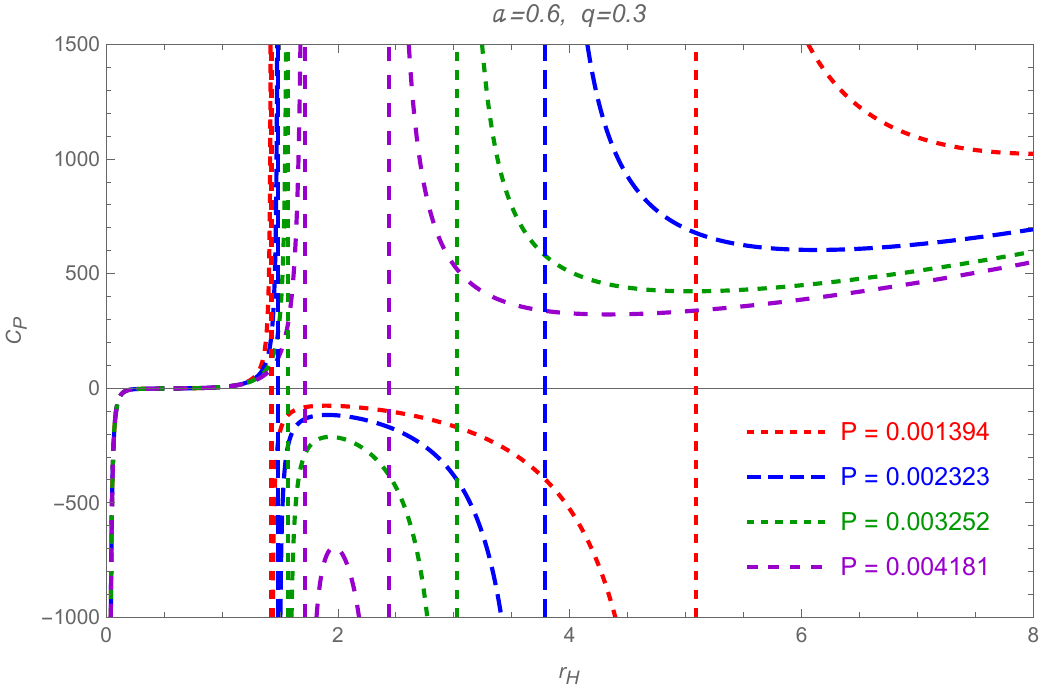}}\hspace{0cm}
\subfloat[][]{\includegraphics[scale=0.47]{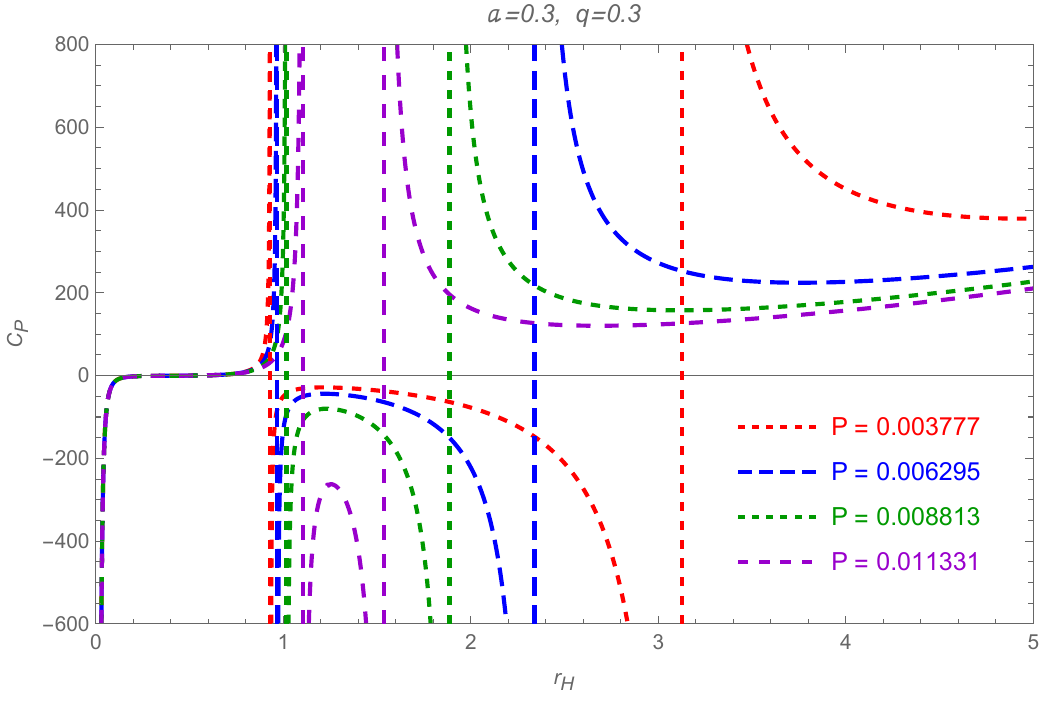}}\\[2mm]
\subfloat[][]{\includegraphics[scale=0.47]{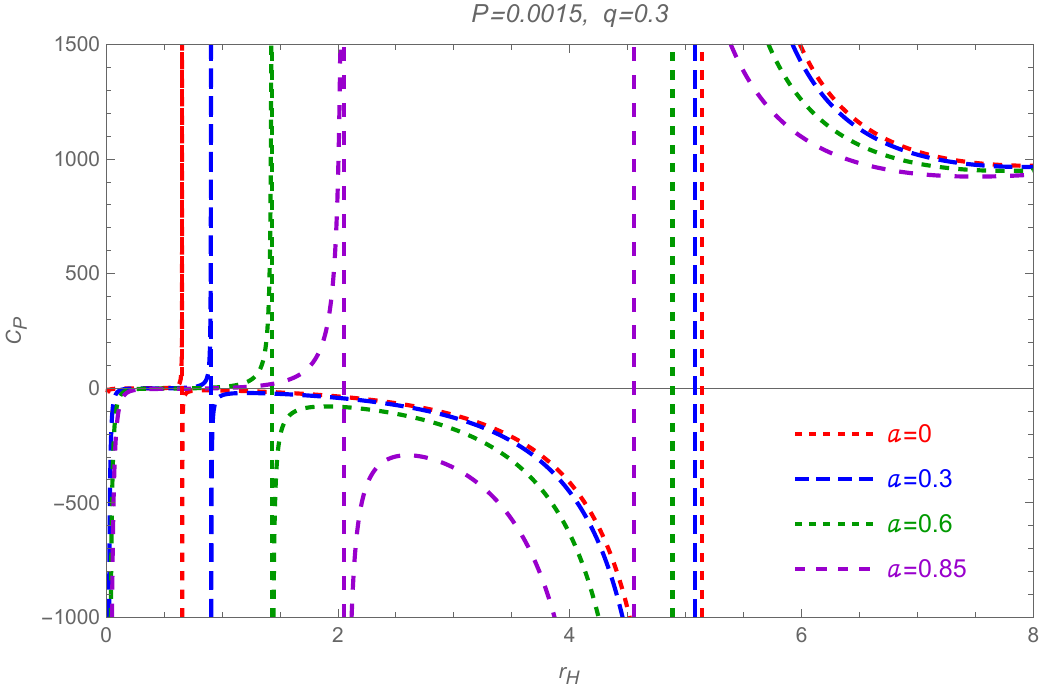}}\hspace{0cm}
\subfloat[][]{\includegraphics[scale=0.47]{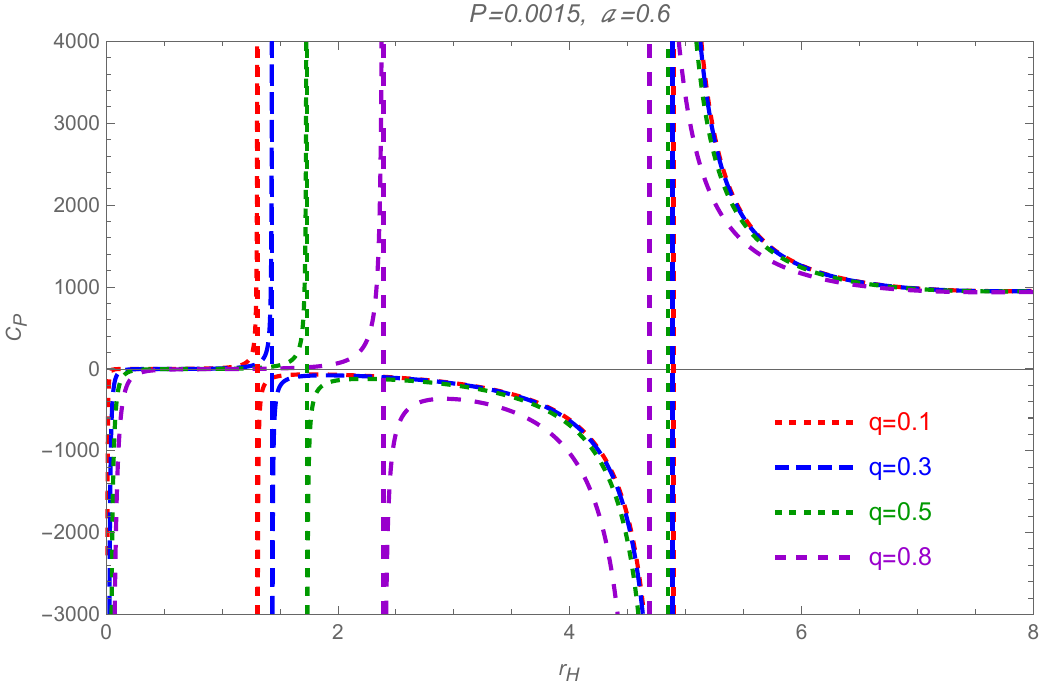}}
\caption{The behavior of $C_{P2}$ vs $r_H$ at $P<P_c$ for different parameters $a$, $q$, and $P$.}
\label{fig:C-P}
\end{figure*}

\begin{figure*}
\centering
\subfloat[][]{\includegraphics[scale=0.47]{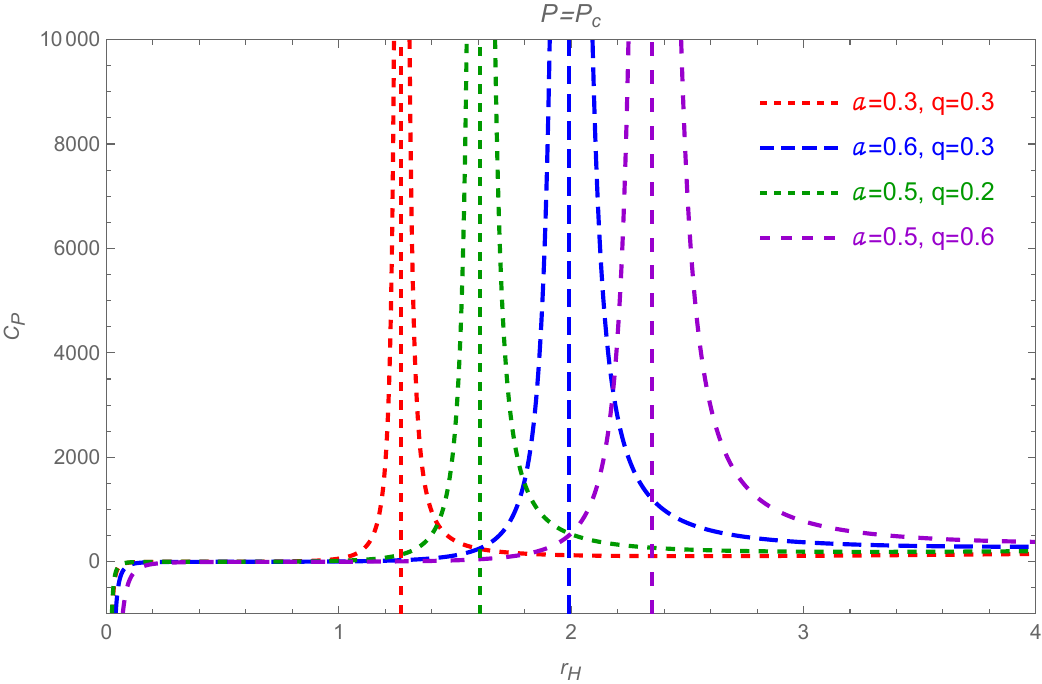}}\hspace{0cm}
\subfloat[][]{\includegraphics[scale=0.47]{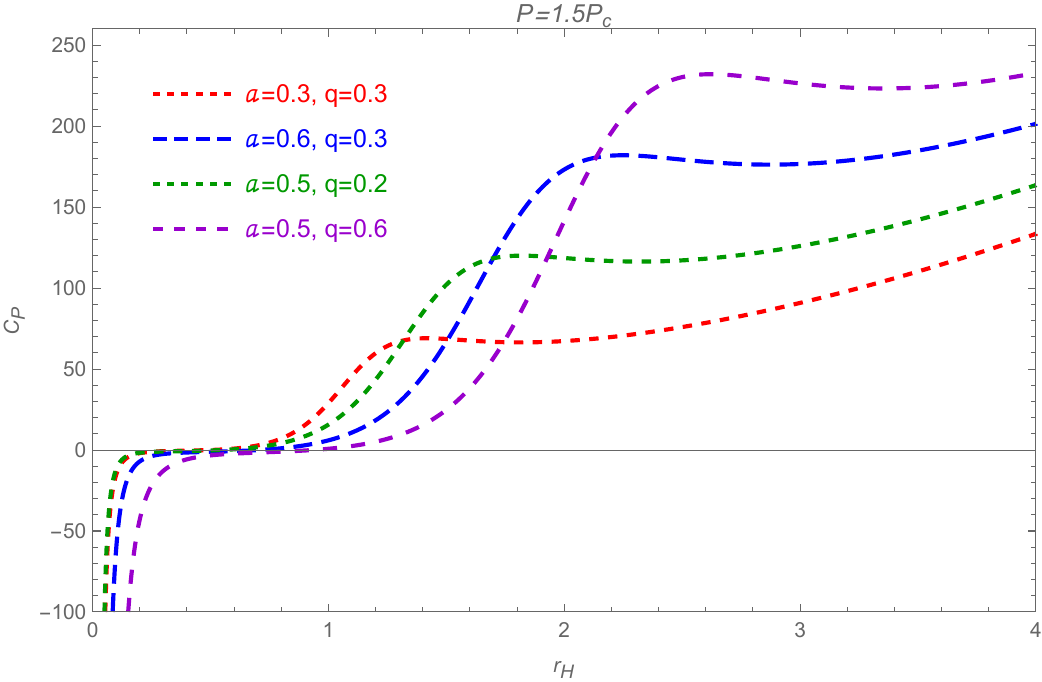}}
\caption{The behavior of $C_{P2}$ vs $r_H$ at $P=P_c$ and $P>P_c$ for different parameters $a$ and $q$.}
\label{fig:C-Pcrit}
\end{figure*}

In this section, we investigate the thermodynamic stability and phase structure of the rotating Hayward-AdS black hole. Because the regularity condition $M=q^3/\alpha$ constrains the black hole parameters, the order in which the constraint is applied remains essential. The two categories of thermodynamic quantities must be obtained in different ways. The fundamental conjugate quantities, such as $T_2$, $\Omega$, $\Phi_2$, $\mathcal{A}_2$, and $V_2$, must be obtained by first differentiating in the unconstrained ``mother'' phase space and then imposing the constraint. The thermodynamic response functions of the regular black hole, such as the Gibbs free energy $G_2$ and the heat capacity $C_{P2}$, must instead be constructed directly from the constrained fundamental quantities. This distinction is especially important for the heat capacity. Imposing the regularity condition on the heat capacity $C_{P1}$ of the ``mother'' black hole does not give the heat capacity $C_{P2}$ of the regular black hole.

\subsection{Gibbs free energy and phase transitions}
\label{sec:4.1}

At fixed pressure, the globally preferred phase of the black hole is determined by the Gibbs free energy. The Gibbs free energy is defined by the Legendre transformation $G = M - TS - \Omega J$. For the rotating Hayward-AdS black hole, it must be constructed from the constrained thermodynamic variables. Substituting Eqs.~(\ref{eq:31}), (\ref{eq:32}), (\ref{eq:38}), and (\ref{eq:39}) into the definition of the Gibbs free energy, we obtain
\begin{equation}\label{eq:41}
\begin{aligned}
G_2 &= M_2 - T_2 S - \Omega J\\
&= \frac{1}{4(-3+8\pi P a^{2})\, r_H^{2}\, (q^{3}+r_H^{3})}
   \Bigl[ - 2q^{6}(3+8\pi P r_H^{2})\\
&\quad + r_H^{4}(a^{2}+r_H^{2})(-3+8\pi P r_H^{2})\\
&\quad - 2q^{3}\bigl(r_H^{3}(9+16\pi P r_H^{2})+a^{2}r_H(6+8\pi P r_H^{2})\bigr)\Bigr].
\end{aligned}
\end{equation}

From the above equation and Eq.~(\ref{eq:39}), we obtain the behavior of $G_2$ with respect to $T_2$ for different parameters $a$, $q$, and $P$, as shown in Fig.~\ref{fig:GT_3curves}. For $P<P_c$, the $G_2$--$T_2$ curve exhibits an 8-shaped or a C-shaped structure, which differs from the standard swallowtail of the RN-AdS and KN-AdS black holes~\cite{Yang:2023,Cheng:2016KNA}. At $P=P_c$, a critical inflection point marks a second-order phase transition. For $P>P_c$, the curve increases monotonically and no phase transition exists.

To study the phase structure of this black hole in more detail, we take $a=0.6$, $q=0.3$ as an example, for which $P_c=0.004646$. In addition to $P_c$, there exists a characteristic pressure $P_z=0.00318<P_c$. The behavior of $G_2$ with respect to $T_2$ for $a=0.6$, $q=0.3$, and different pressures $P$ is shown in Fig.~\ref{fig:GT_4panels}. When $P<P_z$, the curve closes into an 8-shaped structure whose self-intersection point marks a first-order SBH--LBH phase transition (Fig.~\ref{fig:GT_4panels}b). When $P\ll P_z$, the upper loop is small and the lower loop is large, giving the 8-shape a swallowtail-like appearance (Fig.~\ref{fig:GT_4panels}a). As $P$ increases towards $P_z$, the upper loop enlarges while the lower loop shrinks. At $P=P_z$, the lower loop disappears and the curve becomes a single-loop zero-like shape (Fig.~\ref{fig:GT_4panels}c). When $P_z<P<P_c$, the curve becomes a C-shaped structure (Fig.~\ref{fig:GT_4panels}d). In this regime, the Gibbs free energy undergoes a discontinuous jump, signaling a zeroth-order phase transition. The above analysis shows that the rotating Hayward-AdS black hole possesses a richer phase structure than the standard Kerr-AdS and KN-AdS black holes, involving both first-order and zeroth-order phase transitions, which originates from the constraint-induced modifications in the thermodynamic phase space~\cite{Ma:2025bardeen,Xia:2026nsp}.

\subsection{Heat capacity and local thermodynamic stability}
\label{sec:4.2}
Local thermodynamic stability of the rotating Hayward-AdS black hole is governed by the heat capacity at fixed pressure, rotation parameter, and charge,
\begin{equation}\label{eq:43}
C_{P2} \equiv \left(\frac{dM_2}{dT_2}\right)_{P,a,q},
\end{equation}
where $C_{P2}>0$ ($C_{P2}<0$) denotes a locally stable (unstable) configuration. The divergences of $C_{P2}$ occur at the stationary points of $T_2$ discussed in Sec.~\ref{sec:3}, where $dT_2/dr_H=0$ and $dM_2/dr_H\neq0$. Substituting Eqs.~(\ref{eq:38}) and (\ref{eq:39}) into the above equation, we obtain
\begin{equation}\label{eq:CPexplicit}
\begin{aligned}
C_{P2} = \dfrac{18\pi (a^{2}+r_H^{2})^{2}(q^{3}+r_H^{3})^{2}\,
\mathcal{N}_{T}}
{(3-8\pi P a^{2})^{2}\, r_H^{3}\,
\mathcal{D}},
\end{aligned}
\end{equation}
where
\begin{align*}
\mathcal{N}_{T} &= 3\bigl(-2q^{3}r_H^{2}+r_H^{5}+8\pi P r_H^{7}\bigr)\\
&\quad - a^{2}\bigl(3r_H^{3}-8\pi P r_H^{5}+4q^{3}(3+4\pi P r_H^{2})\bigr),\\
\mathcal{D} &= 2a^{2}r_H^{2}\bigl(6r_H^{6}+32\pi P r_H^{8}\bigr)\\
&\quad + 2a^{2}r_H^{2}\bigl(16q^{3}r_H^{3}(3+7\pi P r_H^{2})+q^{6}(15+8\pi P r_H^{2})\bigr)\\
&\quad + 3r_H^{4}\bigl(2q^{6}+r_H^{6}(-1+8\pi P r_H^{2})+2q^{3}r_H^{3}(5+16\pi P r_H^{2})\bigr)\\
&\quad + a^{4}\bigl(-4q^{6}(-3+4\pi P r_H^{2})\bigr)\\
&\quad + a^{4}\bigl(r_H^{6}(3+8\pi P r_H^{2})+q^{3}(42r_H^{3}+64\pi P r_H^{5})\bigr).
\end{align*}
The explicit expression for $C_{P2}$ shows that its divergences, determined by the denominator $\mathcal{D}=0$, coincide with the stationary points of $T_2$, in agreement with the discussion in Sec.~\ref{sec:3}.

The behavior of the heat capacity $C_{P2}$ with respect to $r_H$ for $P<P_c$ and $P\geq P_c$ is shown in Fig.~\ref{fig:C-P} and Fig.~\ref{fig:C-Pcrit}, respectively. The divergence structure of $C_{P2}$ follows directly from the stationary points of the $T_2$--$r_H$ curve. For $P<P_c$, $T_2$ has a local maximum at smaller $r_H$ and a local minimum at larger $r_H$, and $C_{P2}$ accordingly develops two divergences that divide the black holes into the SBH, IBH, and LBH branches. The SBH and LBH branches have $C_{P2}>0$ and are locally stable, whereas the IBH branch has $C_{P2}<0$ and is locally unstable. The two divergences at $r_{H1}$ and $r_{H2}$ mark two second-order phase transitions, namely the SBH--IBH transition at $r_{H1}$ and the IBH--LBH transition at $r_{H2}$.

For $P<P_c$, as shown in Fig.~\ref{fig:C-P}, the region of the IBH branch is reduced as any one of the parameters $P$, $a$, or $q$ is increased. The two divergence points, however, move in different ways for different parameters. Increasing the pressure $P$ (panel a, with $a=0.6$ and $q=0.3$) reduces the region of the IBH branch mainly from the right. For example, when $P$ increases from $0.3P_c=0.001394$ to $0.5P_c=0.002323$ (the red and blue curves in panel a), the right-hand point $r_{H2}$ moves leftward from $5.090$ to $3.783$, whereas the left-hand point $r_{H1}$ only drifts rightward from $1.425$ to $1.485$, so that the IBH width is reduced from $\Delta r_H=3.665$ to $\Delta r_H=2.297$. In contrast, increasing the rotation parameter $a$ (panel c, with $q=0.3$ and $P=0.0015$ fixed) or the charge $q$ (panel d, with $a=0.6$ and $P=0.0015$ fixed) reduces the region of the IBH branch mainly from the left. For example, as $a$ increases from $0$ to $0.3$ (the red and blue curves in panel c), the left-hand point $r_{H1}$ moves rightward from $0.655$ to $0.905$, whereas the right-hand point $r_{H2}$ only drifts leftward from $5.143$ to $5.083$, remaining nearly fixed. Thus, increasing $P$ reduces the region of the IBH branch from the right, whereas increasing $a$ or $q$ reduces its region from the left.

At $P=P_c$, the two divergence points coalesce into a single one at $r_H=r_c$, marking a second-order phase transition, and $C_{P2}$ is positive, so that the black hole is locally stable. For $P>P_c$, no divergence remains, and $C_{P2}$ is positive as well, so that the black hole is locally stable, as shown in Fig.~\ref{fig:C-Pcrit}. Moreover, in all regimes, whether $P<P_c$, $P=P_c$, or $P>P_c$, the heat capacity remains negative near the origin, so that the very small black hole is always locally thermodynamically unstable.

\subsection{$P$--$V$ criticality}
\label{sec:4.3}
\begin{figure*}
\centering
\subfloat[][]{\includegraphics[scale=0.47]{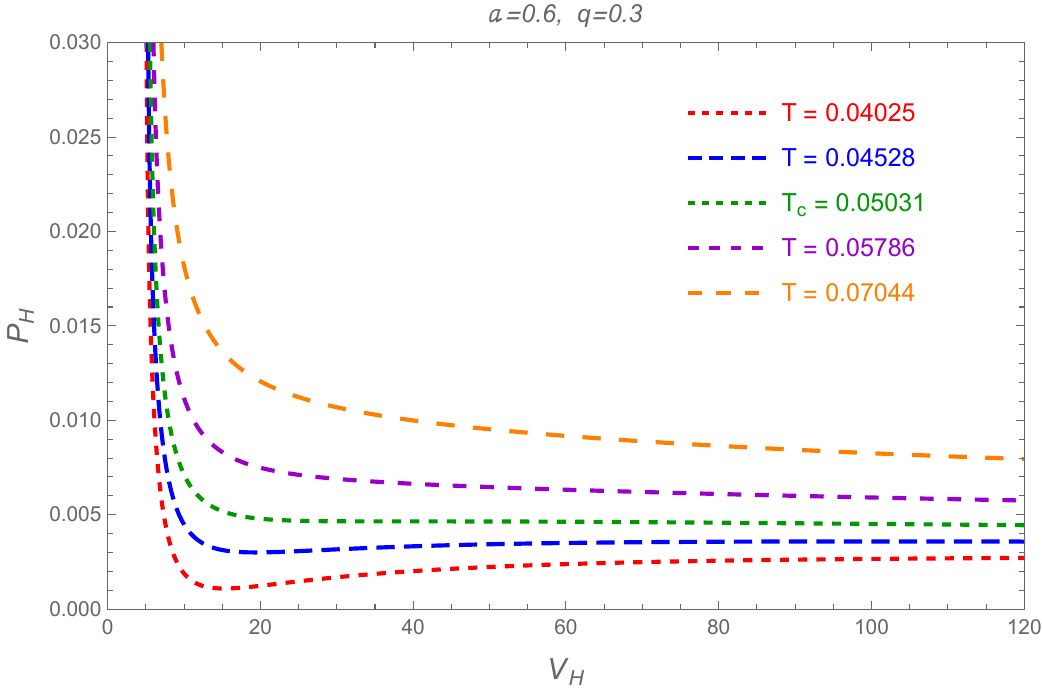}}\hspace{0cm}
\subfloat[][]{\includegraphics[scale=0.47]{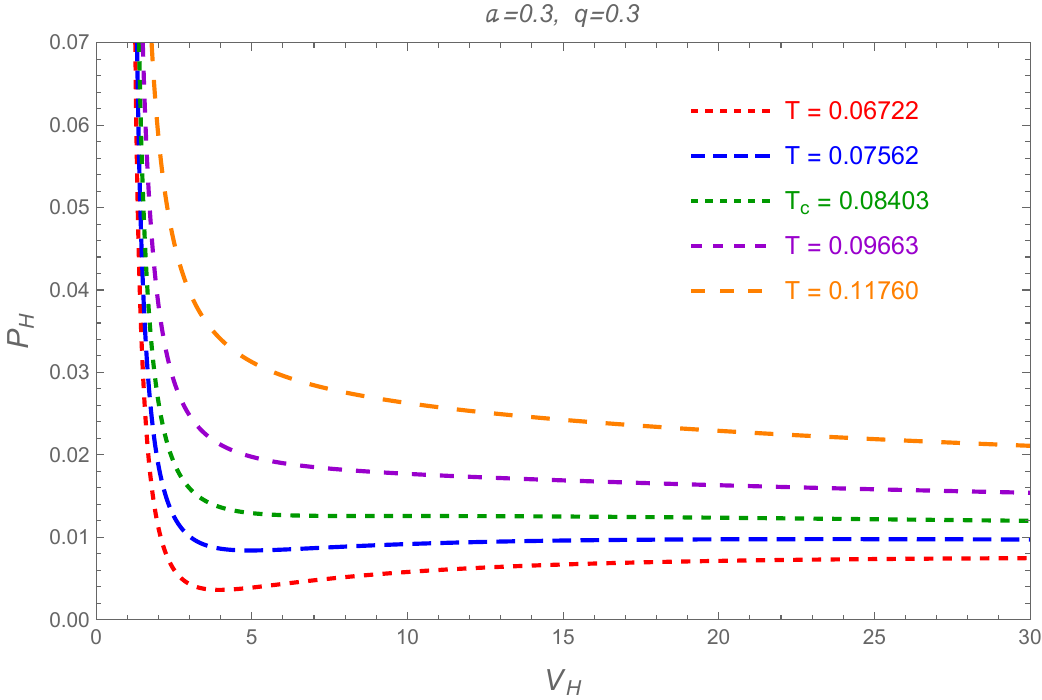}}\\[2mm]
\caption{The behavior of $P$ vs $V$ for different parameters $a$, $q$ and $T$.}
\label{fig:PV}
\end{figure*}
From Eq.~(\ref{eq:39}), we can obtain the relationship between pressure and Hawking temperature,
\begin{equation}\label{eq:PofT}
\begin{aligned}
P &= \frac{1}{8\pi\bigl(3r_H^{7}+a^{2}r_H^{5}-2a^{2}q^{3}r_H^{2}\bigr)}\\
&\quad \times \Bigl[12\pi r_H(r_H^{2}+a^{2})(q^{3}+r_H^{3})\,T_2\\
&\quad - 3\bigl(r_H^{5}-2q^{3}r_H^{2}-a^{2}r_H^{3}-4a^{2}q^{3}\bigr)\Bigr].
\end{aligned}
\end{equation}
Using the above equation and the thermodynamic volume $V_2$ in Eq.~(\ref{eq:312}), the behavior of the pressure $P$ with respect to the thermodynamic volume $V$ for different temperatures $T$ is shown in Fig.~\ref{fig:PV}. In the $P$--$V$ diagram, the critical point corresponds to the temperature at which the local maximum and minimum of the isotherm coalesce into a single inflection point. This point satisfies
\begin{equation}\label{eq:46}
\begin{aligned}
\left(\dfrac{\partial P}{\partial r_H}\right)_{T_2,a,q} = 0,
\qquad
\left(\dfrac{\partial^{2} P}{\partial r_H^{2}}\right)_{T_2,a,q} = 0.
\end{aligned}
\end{equation}
Solving Eqs.~(\ref{eq:PofT}) and (\ref{eq:46}), we obtain the critical temperature and critical pressure as functions of the horizon radius $r_H$ and the parameters $a$ and $q$,
\begin{equation}\label{eq:47}
\begin{aligned}
T_c &= \frac{1}{2\pi r_H\,\mathcal{W}}
     \Bigl(8a^{4}q^{6} - 11a^{4}q^{3}r_H^{3} - a^{4}r_H^{6} - 42a^{2}q^{3}r_H^{5}\\
&\quad  - 6a^{2}r_H^{8}- 15q^{3}r_H^{7} + 3r_H^{10}\Bigr),
\end{aligned}
\end{equation}
\begin{equation}\label{eq:48}
\begin{aligned}
P_c &= -\frac{3}{8\pi r_H^{2}\,\mathcal{W}}
     \Bigl(4a^{4}q^{6} + 14a^{4}q^{3}r_H^{3} + a^{4}r_H^{6} + 10a^{2}q^{6}r_H^{2}\\
&\quad  + 32a^{2}q^{3}r_H^{5} + 4a^{2}r_H^{8}+ 2q^{6}r_H^{4} + 10q^{3}r_H^{7} - r_H^{10}\Bigr),
\end{aligned}
\end{equation}
where $\mathcal{W} = -2a^{4}q^{6} + 8a^{4}q^{3}r_H^{3} + a^{4}r_H^{6} + 2a^{2}q^{6}r_H^{2} + 28a^{2}q^{3}r_H^{5} + 8a^{2}r_H^{8} + 12q^{3}r_H^{7} + 3r_H^{10}$.
For example, for $a=0.6$, $q=0.3$ and $a=0.3$, $q=0.3$, we obtain the critical values $T_c=0.0503$, $P_c=0.004646$ and $T_c=0.0840$, $P_c=0.012590$, respectively. From Fig.~\ref{fig:PV}, the isotherms display three distinct behaviors depending on the temperature. When $T<T_c$ (the red and blue curves), each isotherm possesses an oscillatory segment, with a local minimum followed by a local maximum, which signals an SBH--LBH phase transition of the van der Waals type. The Maxwell equal-area construction selects the physically relevant branch, where the SBH and LBH phases coexist. When $T=T_c$ (the green curve), the oscillation shrinks and the two stationary points merge into an inflection point, where a second-order phase transition occurs. When $T>T_c$ (the purple and orange curves), the isotherms become monotonic and approach the behavior of an ideal gas.

\section{Conclusions and discussion}
\label{sec:5}

We have systematically studied the thermodynamics, phase transitions, and local stability of the rotating Hayward-AdS regular black hole in the extended phase space $(S, J, q, \alpha, P)$. Starting from the spherically symmetric ``mother'' black hole of the Hayward-AdS black hole, we have obtained its rotating ``mother'' black hole metric using the modified Newman--Janis algorithm, and have returned to the rotating Hayward-AdS black hole solution by imposing the regularity condition.

We have derived the thermodynamic quantities of the rotating regular black hole in two steps. We have first obtained the fundamental thermodynamic quantities of the singular ``mother'' black hole from the first law of black hole thermodynamics in the unconstrained phase space $(S, J, q, \alpha, P)$, and have then imposed the regularity condition $M = q^{3}/\alpha$ on them, thereby obtaining the fundamental thermodynamic quantities of the rotating Hayward-AdS regular black hole. This two-step procedure is essential, because the regularity condition restructures the thermodynamic phase space. Accordingly, the two categories of thermodynamic quantities must be obtained in different ways: the fundamental conjugate quantities $T_2$, $\Omega$, $\Phi_2$, $\mathcal{A}_2$, and $V_2$ must be obtained by first differentiating and then imposing the constraint, whereas the thermodynamic response functions, such as the Gibbs free energy $G_2$ and the heat capacity $C_{P2}$, must be constructed directly from the constrained fundamental quantities. The resulting temperature is consistent with the surface gravity, thereby providing a self-consistent thermodynamics for the rotating Hayward-AdS regular black hole.

We have explored the phase structure through the $G_2$--$T_2$ diagram and the $P$--$V$ isotherms. The global phase structure is governed by the critical pressure $P_c$ and the characteristic pressure $P_z<P_c$. For $P<P_c$, the system exhibits three branches, namely SBH, IBH, and LBH, and the $P$--$V$ isotherms display the oscillatory behavior characteristic of a liquid/gas phase transition. The pressure $P_z$ divides the region below the critical pressure into two different parts. For $P<P_z$, the $G_2$--$T_2$ curve forms an 8-shaped structure, whose self-intersection marks a first-order SBH--LBH phase transition. For $P_z<P<P_c$, the curve becomes C-shaped, corresponding to a zeroth-order SBH--LBH phase transition in which the Gibbs free energy undergoes a discontinuous jump. At $P=P_c$, a second-order phase transition exists, while for $P>P_c$ no phase transition occurs.

The local thermodynamic stability is determined by the heat capacity $C_{P2}$. For $P<P_c$, the Hawking temperature $T_2$ possesses a local maximum and a local minimum, and $C_{P2}$ correspondingly develops two divergence points, which divide the black holes into three branches. The SBH and LBH branches are locally stable for $C_{P2}>0$, whereas the IBH branch is locally unstable for $C_{P2}<0$. At $P=P_c$, the two stationary points of $T_2$ merge and the two divergence points of $C_{P2}$ coalesce accordingly, and $C_{P2}$ is positive, so that the black hole is locally stable. For $P>P_c$, $C_{P2}$ has no divergence points and remains positive, so that the black hole is locally stable and no phase transition occurs. In addition, $C_{P2}$ remains negative near the origin for all pressures, indicating that very small regular black holes are always locally thermodynamically unstable. As any one of the parameters $P$, $a$, or $q$ is increased, the region of the IBH branch is reduced, but in different ways. Increasing $P$ moves the right-hand point $r_{H2}$ leftward and reduces the region of the IBH branch mainly from the right, whereas increasing $a$ (or $q$) moves the left-hand point $r_{H1}$ rightward while $r_{H2}$ remains almost unchanged, thus reducing the region of the IBH branch mainly from the left.

In contrast to the RN-AdS and Kerr--Newman-AdS black holes, which exhibit the standard swallowtail free energy, the rotating Hayward-AdS black hole displays 8-shaped and C-shaped phase structures. This indicates that the non-standard phase structure stems from the regularity constraint rather than from spherical symmetry, and that rotation does not restore the swallowtail. We expect that other rotating regular black holes generated by nonlinear electrodynamics sources exhibit similar behavior. More generally, the regularity condition restructures the thermodynamic phase space and thereby reshapes the phase transitions themselves.

\subsection*{Acknowledgments}
The work is supported by the National Natural Science Foundation of China (Grant No. 12463001), and the Yunnan Xingdian Talent Support Program - Young Talent Project.

\end{document}